\documentclass[%
reprint,
showpacs,
 amsmath,amssymb,
 aps,
pra,
]{revtex4-1}

\newcommand\beq{\begin{equation}}
\newcommand\eeq{\end{equation}}

\usepackage{graphicx}% Include figure files
\usepackage{dcolumn}% Align table columns on decimal point
\usepackage{bm}% bold math
\begin{document}

%\preprint{APS/123-QED}

\title{Non-Hermiticity-Induced Wave Confinement and Guiding\\
in Loss-Gain-Loss Three-Layer Systems}

\author{Silvio Savoia}
\author{Giuseppe Castaldi}
\author{Vincenzo Galdi}
\email{vgaldi@unisannio.it}
\affiliation{Waves Group, Department of Engineering, University of Sannio, I-82100 Benevento, Italy
}%

\date{\today}% It is always \today, today,

%%%%%%%%%%%%%%%%%%%% Created 14/07/2016
%%%%%%%%%%%%%%%%%%%% Last revised 27/09/2016

\begin{abstract}
Following up on previous studies on {\em parity-time-symmetric} gain-loss bi-layers, and inspired by formal analogies with plasmonic waveguides, we study non-Hermiticity-induced wave confinement and guiding phenomena that can occur in loss-gain-loss three-layers. By revisiting previous well-established ``gain-guiding'' concepts, we investigate analytically and numerically the dispersion and confinement properties of guided modes that can be supported by this type of structures, by assuming realistic dispersion models and parameters for the material constituents. As key outcomes, we identify certain modes with specific polarization and symmetry that exhibit particularly desirable characteristics, in terms of quasi-real propagation constant and sub-wavelength confinement. Moreover, we
elucidate the effects of material dispersion and parameters, and highlight the potential advantages by comparison with the previously studied gain-loss bi-layer configurations. Our results provide additional perspectives on light control in non-Hermitian optical systems, and may find potentially intriguing applicability to reconfigurable nanophotonic platforms.
\end{abstract}

\maketitle

%%%%%%%%%%%%%%%%%%%%%%%%%%%%%%%%%%%%%%%%%%%%%%%%%%%%%%%%%%%%%%%%%%
\section{Introduction}
%%%%%%%%%%%%%%%%%%%%%%%%%%%%%%%%%%%%%%%%%%%%%%%%%%%%%%%%%%%%%%%%%%

The concepts of {\em parity-time} ($\mathcal{PT}$) symmetry and {\em non-Hermitian} extensions of quantum mechanics, introduced in a series of seminal studies by Bender and co-workers \cite{Bender:1998,Bender:2002,Bender:2007}, have resonated in several research communities, triggering a surge of interest in the study of non-Hermitian systems. In optics and photonics, where non-Hermiticity is associated with the presence of loss and/or gain, these ideas have inspired novel, {\em unconventional} ways of mixing material constituents featuring loss and gain, so as to attain a wealth of anomalous light-matter interactions far more sophisticated that mere loss-compensation effects (see, e.g., \cite{Zyablovsky:2014rf} for a recent review of $\mathcal{PT}$-symmetry in optics).

During the past few years, the emerging field of ``non-Hermitian optics'' has gained a steadily growing attention in both basic and applied research. For instance, of crucial interest for basic physics is the possibility to design optical analogues of quantum-physics scenarios, so as to experimentally test some controversial implications of non-Hermitian quantum field theories \cite{Longhi:2012pr}, as well as to gain a deeper understanding of phenomena and properties that are typical of non-Hermitian systems, such as spontaneous symmetry breaking \cite{Ruter:2010} and exceptional points \cite{Longhi:2014km,Zhen:2015sp,Hahn:2016dv,Cerjan:2016bv}, unidirectional invisibility \cite{Lin:2011ui,Regensburger:2012}, Bloch oscillations \cite{Xu:2016bh}, and
coherent perfect absorption \cite{Longhi:2010pt,Chong:2011ev,Sun:2014:lk}, just to mention a few.

From the application viewpoint, a broad variety of effects and configurations have been proposed and explored, ranging from lasers to metamaterials (see \cite{Ctyroky:2010kl,Benisty:2011jr,Schindler:2011es,Lazarides:2013gd,Zhu:2013jf,Castaldi:2013pt,Kulishov:2013gs,Ming:2013es,Makris:2014rg,Alaeian:2014eb,Alaeian:2014dj,Savoia:2014,Peng:2014,Peng:2014bc,Fleury:2014nr,Silveirinha:2014sp,Sounas:2015uc,Principe:2015fw,Savoia:2015,Alaeian:2015df,Benisty:2015gb,Longhi:2015kj,Poli:2015bc,Savoia:2016,Radi:2016gf,Chen:2016li,Hurwitz:206hg} and references therein).
In particular, although the general concept of ``gain guiding'' is well-established in linear \cite{Siegman:2003pm} and nonlinear \cite{Zezyulin:2011si} optics, the $\mathcal{PT}$-symmetry concept has inspired additional perspectives in
wave confinement and guiding. For instance,
as shown in \cite{Ctyroky:2010kl}, a $\mathcal{PT}$-symmetric gain-loss (GL) half-space or bi-layer is capable to sustain surface waves whose dispersion equation formally resembles that of surface-plasmon-polaritons (SPPs) \cite{Maier:2007kw}. These surface waves propagate unattenuated along the gain-loss interface, and are transversely confined with exponential decay controlled by the gain/loss level. In \cite{Savoia:2015}, we showed that in the (real-part) {\em epsilon-near-zero} regime 
the gain/loss level necessary to sustain these effects can be significantly reduced. 

The above waveguiding mechanism opens up potentially intriguing perspectives in reconfigurable nanophotonic platforms. For instance, one can envision deploying ``channels'' made of gain media in a lossy background, which could be selectively activated via optical pumping. Within this framework, it remains an open question to what extent this mechanism is effective when realistic material dispersion models are taken into account \cite{Zyablovsky:2014cp}, and whether there is room for potential improvements. In fact, building up on the above-mentioned SPP analogy, one might wonder whether {\em three-layer} non-Hermitian configurations (analogous to the insulator-metal-insulator \cite{Goto:2004fh,Charbonneau:2005bp} and metal-insulator-metal \cite{Zia:2004as,Dionne:2006tg} heterostructures) could potentially offer any advantages.

To answer this question, in this paper, we revisit the  wave confinement and guiding effects induced by non-Hermiticity in a {\em loss-gain-loss} (LGL) three-layer geometry. By comparison with previous studies \cite{Siegman:2003pm}, our results are not restricted to the weakly-guiding regime, incorporate realistic material dispersion models, and focus on the conditions to attain well-confined modes with quasi-real propagation constants.
More in detail, in Sec. \ref{Sec:Statement}, we outline the problem statement, with the relevant geometry, parameters and mathematical formulation. Subsequently, in Sec. \ref{Sec:Results}, we discuss some representative numerical results from our parametric studies, assuming realistic Lorentz-type dispersion models for the gain and loss media. Within this framework, we illustrate the dispersion properties of the fundamental transverse-magnetic (TM) and transverse-electric (TE) modes supported by LGL three-layers, highlighting the non-Hermiticity-induced effects as well as the similarities, differences and potential advantages by comparison with the previously studied $\mathcal{PT}$-symmetric GL bi-layer case. Moreover, we elucidate the effects of material dispersion and gain/loss level.
Finally, some brief concluding remarks follow in Sec. \ref{Sec:Conclusions}, while Appendices \ref{Sec:AppendixA} and \ref{Sec:AppendixB}
contain some ancillary details on the modal solutions and numerical simulations.

%%%%%%%%%%%%%%%%%%%%%%%%%%%%%%%%%%%%%%%%%%%%%%%%%%%%%%%%%%%%%%%%%%
\section{Problem Statement}
%%%%%%%%%%%%%%%%%%%%%%%%%%%%%%%%%%%%%%%%%%%%%%%%%%%%%%%%%%%%%%%%%%
\label{Sec:Statement}

%############################################################
%                Figure1
%
\begin{figure}
\begin{center}
\includegraphics [width=8cm]{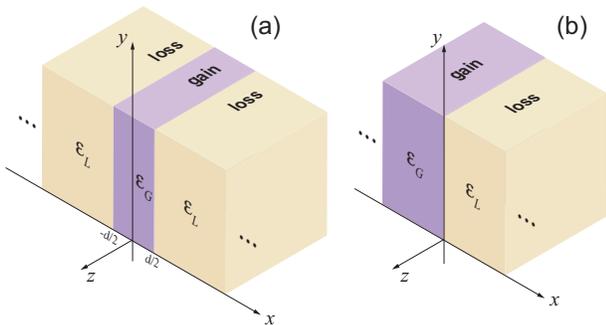}% Here is how to import EPS art
\end{center}
\caption{(Color online) Problem schematic. (a) LGL three-layer configuration, consisting of a layer of gain medium sandwiched between two lossy half-spaces, with relative permittivity distribution as in (\ref{eq:LGL}). (b) GL bi-layer configuration, with relative permittivity distribution as in (\ref{eq:GL}). All field quantities are assumed as $y$-independent, and propagation is studied along the $z$-direction.}
\label{Figure1}
\end{figure}
%############################################################

%-----------------------------------------------------------------
\subsection{Geometry, Parameters and Generalities}
%-----------------------------------------------------------------
Referring to the schematic in Fig. \ref{Figure1}(a), we consider a three-layer LGL configuration comprising a gain-medium layer sandwiched between two lossy half-spaces, mathematically described (in the assumed reference system) by the relative permittivity distribution  
\beq
\varepsilon_{LGL}\left(x,\omega\right)=\left\{
\begin{array}{ll}
\varepsilon_L\left(\omega\right),~~\left|x\right|>d/2,\\
\varepsilon_G\left(\omega\right),~~\left|x\right|<d/2,
\label{eq:LGL}
\end{array}
\right.
\eeq
with $\varepsilon_L$ and $\varepsilon_G$ denoting the relative permittivities of the loss and gain media, respectively, and $d$ the thickness of the gain-medium layer.  Although, as previously mentioned, such three-layer configuration is somehow inspired by plasmonic heterostructures, the analogy is limited to the formal structure of the dispersion equations, and there is no meaningful physical correspondence between gain/loss and insulator/metal materials. However, the choice of an LGL (rather than gain-loss-gain) arrangement allows a straightforward application of the radiation condition and decay at infinity (see Appendix \ref{Sec:AppendixA} for details), whereas it is well-known that semi-infinite gain layers would lead to {\em unstable} solutions (see the discussion in Sec. 2.B in \cite{Siegman:2003pm}). 

Also of interest, for comparison purposes, is the GL bi-layer configuration shown in Fig. \ref{Figure1}(b),
\beq
\varepsilon_{GL}\left(x,\omega\right)=\left\{
\begin{array}{ll}
\varepsilon_G\left(\omega\right),~~x<0,\\
\varepsilon_L\left(\omega\right),~~x>0,
\label{eq:GL}
\end{array}
\right.
\eeq
which was studied in \cite{Ctyroky:2010kl,Savoia:2015}.

In what follows, the materials are assumed as non-magnetic (relative permeability $\mu=1$), and the regions are assumed as uniform and of infinite extent along the $y$- and $z$-directions, so as to reduce the problem to a two-dimensional form. To further simplify the analytical treatment, we also assume unbounded half-spaces along the $x$-direction (see Sec. \ref{Sec:Remarks} below for a brief discussion of the related truncation effects). 

We are interested in studying the time-harmonic [$\exp(-i\omega t)$] wave propagation along the $z$-direction, with all observables assumed as independent of $y$. As it was recently pointed out in connection with optical $\mathcal{PT}$-symmetry \cite{Zyablovsky:2014cp}, it is important to account for material dispersion in the study of non-Hermitian optical systems. Accordingly, for the loss and gain media, we assume realistic Lorentz-type dispersion models, 
\begin{subequations}
\begin{eqnarray}
{\varepsilon_L}\left(\omega\right) &=& \varepsilon'_0 - \frac{\Gamma \varepsilon''_0\omega_0}
{\omega^2-\omega_0^2+i\Gamma \omega},\\
\label{eq:eps_L}
{\varepsilon_G}\left(\omega\right) &=& \varepsilon'_0 + \frac{\Gamma \varepsilon''_0\omega_0}
{\omega^2-\omega_0^2+i\Gamma \omega},
\label{eq:eps_G}
\end{eqnarray}
\label{eq:Lorentz}
\end{subequations}
where $\omega_0$ denotes the chosen operational radian frequency (henceforth simply referred to as ``center frequency''), 
$\varepsilon'_0>0$ represents the high-frequency limit (as well as the real-part at $\omega=\omega_0$),  $\varepsilon''_0>0$ is the peak gain/loss level (occurring at $\omega=\omega_0$), and $\Gamma$ is a dampening factor that controls the resonance lineshape. Here, and henceforth, the subscript ``0'' is used to identify quantities evaluated at the center frequency. 

The above models are physically consistent with typical processes underlying the tailoring of loss and gain at optical frequencies, e.g., the introduction of absorptive or active dopants such as two-level atoms (or quantum dots). Albeit not strictly necessary, in order to minimize the number of relevant parameters, we choose identical values for $\varepsilon'_0$, $\varepsilon''_0$ and $\Gamma$ in the gain and loss media. One consequence of this choice is that, at the center frequency, the permittivities of the gain and loss media are {\em complex conjugate},
\beq
\varepsilon_G\left(\omega_0\right)=\varepsilon^*_L\left(\omega_0\right)=\varepsilon'_0-i\varepsilon''_0,
\label{eq:cc}
\eeq
which, in turn, implies that the GL bi-layer configuration in Fig. \ref{Figure1}(b) satisfies the $\mathcal{PT}$-symmetry condition
\beq
\varepsilon_{GL}\left(-x,\omega_0\right)=\varepsilon_{GL}^*\left(x,\omega_0\right)
\label{eq:PTS}
\eeq
assumed in the previous studies \cite{Ctyroky:2010kl,Savoia:2015}.

%############################################################
%                Figure2
%
\begin{figure}
\begin{center}
\includegraphics [width=8cm]{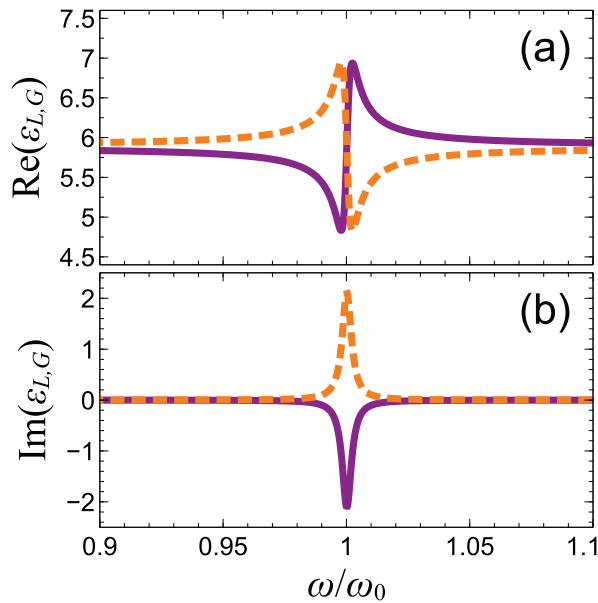}% Here is how to import EPS art
\end{center}
\caption{(Color online) (a), (b) Real and imaginary parts, respectively, of the relative permittivity of the lossy (orange-dashed curves) and gain (purple-solid curves) media, as a function of frequency, assuming the Lorentz-type dispersion models in (\ref{eq:Lorentz}) with $\varepsilon'_0=5.887$, $\varepsilon''_0=2.110$, and $\Gamma=4.523\cdot10^{-3}\omega_0$.}
\label{Figure2}
\end{figure}
%############################################################

Figure \ref{Figure2} shows the real and imaginary parts of the relative permittivities in (\ref{eq:Lorentz}), around the center frequency, for a set of parameters (given in the caption) that are consistent with those utilized in the recent topical literature \cite{Liberal:2014md,Xie:2015af} in connection with gain media based on quantum dots \cite{Campbell:2012cj,Moreels:2012td}. More specifically, the parameters assumed in this example pertain to a commercially available gain medium based on CdSe/ZnS core-shell quantum-dots \cite{Campbell:2012cj} which, at the center wavelength $\lambda_0=560$nm, exhibits a gain coefficient 
\beq
\gamma_0=-\frac{4\pi\mbox{Im}\left[\sqrt{\varepsilon_G\left(\omega_0\right)}\right]}{\lambda_0}\approx 0.96\cdot 10^5\mbox{cm}^{-1}.
\eeq

It is evident from Fig. \ref{Figure2} that the complex-conjugate condition in (\ref{eq:cc}) [and hence the $\mathcal{PT}$-symmetry condition in (\ref{eq:PTS}) for the GL bi-layer in Fig. \ref{Figure1}(b)] is strictly verified {\em only} at $\omega=\omega_0$ since, for different frequencies, the real-parts of $\varepsilon_L$ and $\varepsilon_G$ are different. We also stress that the LGL configuration in Fig. \ref{Figure1}(a) {\em never} satisfies the $\mathcal{PT}$-symmetry condition, irrespective of frequency.

%-----------------------------------------------------------------
\subsection{Dispersion Equations}
%-----------------------------------------------------------------
\label{eq:DE}
Before introducing the mathematical modeling, by simple inspection of the material dispersion laws in Fig. \ref{Figure2}, we can qualitatively anticipate the presence of three frequency regimes for the LGL three-layer configuration:
\begin{enumerate}
\item{For $\omega\lesssim \omega_0$, both imaginary parts are negligible, whereas $\mbox{Re}\left(\varepsilon_L\right)\gtrsim \mbox{Re}\left(\varepsilon_G\right)$. Essentially, this corresponds to an {\em anti-guiding} regime where no guided modes should be expected.}
\item{For $\omega\approx \omega_0$, there is a strong contrast in the imaginary parts, while the real parts tend to be similar. This represents the most interesting region, where non-Hermiticity plays a key role.}
\item{For $\omega\gtrsim\omega_0$, both imaginary parts are once again negligible, but now $\mbox{Re}\left(\varepsilon_G\right)\gtrsim \mbox{Re}\left(\varepsilon_L\right)$. This is basically the operational regime of {\em weakly-guiding} dielectric waveguides.}
\end{enumerate}

%=================================================================
\subsubsection{TM Modes}
%=================================================================
We start considering TM modes, characterized by $y$-directed magnetic field, and electric field laying in the $x-z$ plane. By assuming propagation along the $z$-direction, and exponential decay for $|x|\rightarrow\infty$, the dispersion equation can be straighfowardly derived in formal analogy with the plasmonic case (see, e.g., Sec. 2.3 in \cite{Maier:2007kw} and Appendix \ref{Sec:AppendixA} for details). In particular, following the same convention as in \cite{Maier:2007kw}, we identify modes with {\em even} and {\em odd} vector parity, which satisfy the dispersion equations
\begin{subequations}
\beq
\frac{k_{xL}}{\varepsilon_L\left(\omega\right)}+\frac{i k_{xG}\cot\left(\displaystyle{\frac{k_{xG}d}{2}}\right)}{\varepsilon_G\left(\omega\right)}=0
\label{eq:TM_even}
\eeq
and
\beq
-\frac{k_{xL}}{\varepsilon_L\left(\omega\right)}+\frac{i k_{xG}\tan\left(\displaystyle{\frac{k_{xG}d}{2}}\right)}{\varepsilon_G\left(\omega\right)}=0,
\label{eq:TM_odd}
\eeq
\label{eq:TM}
\end{subequations}
respectively. In (\ref{eq:TM}), 
\begin{subequations}
\begin{eqnarray}
k_{xL}&=&\sqrt{k^2 \varepsilon_L\left(\omega\right)-k_z^2},~~\mbox{Im}\left(k_{xL}\right)\ge 0,
\label{eq:kxL}\\
k_{xG}&=&\sqrt{k^2 \varepsilon_G\left(\omega\right)-k_z^2},
\label{eq:kxG}
\end{eqnarray}
\label{eq:kxLG}
\end{subequations}
indicate the $x$-domain wavenumbers in the loss and gain regions, respectively, with $k_z$ denoting the propagation constant along the $z$-direction, and $k=\omega/c=2\pi/\lambda$ the vacuum wavenumber (with $c$ and $\lambda$ being the corresponding wavespeed and wavelength, respectively).
We highlight that the branch-cut choice in (\ref{eq:kxL}) is consistent with the correct field decay at infinity in the loss regions, whereas this choice is irrelevant in (\ref{eq:kxG}) in view of the finite thickness of the gain layer.

%=================================================================
\subsubsection{TE Modes}
%=================================================================
Although they are not relevant in the plasmonic counterpart, TE modes (with $y$-directed electric field, and magnetic field laying in the $x-z$ plane) can in fact be supported by the LGL three-layer under study. The following dispersion equations can be derived for the {\em even} and {\em odd} modes, respectively (see Appendix \ref{Sec:AppendixA} for details):
\begin{subequations}
\beq
k_{xL}+i k_{xG}\cot\left(\displaystyle{\frac{k_{xG}d}{2}}\right)=0,
\label{eq:TE_even}
\eeq
\beq
-k_{xL}+i k_{xG}\tan\left(\displaystyle{\frac{k_{xG}d}{2}}\right)=0.
\label{eq:TE_odd}
\eeq
\label{eq:TE}
\end{subequations}

%=================================================================
\subsubsection{Gain-Loss Bi-layer}
%=================================================================
\label{Sec:GL}
For completeness, we also briefly review the main properties of TM surface modes supported by the GL bi-layer in Fig. \ref{Figure1}(b).
The corresponding dispersion equation can written as \cite{Ctyroky:2010kl,Savoia:2015} (see also Appendix \ref{Sec:AppendixB}) 
\beq
\frac{k_{xL}}{\varepsilon_L\left(\omega\right)}=\frac{k_{xG}}{\varepsilon_G\left(\omega\right)},
\label{eq:kzPTimpl}
\eeq
with $k_{xL}$ and $k_{xG}$ defined as in ($\ref{eq:kxLG}$), but now assuming $\mbox{Im}\left(k_{xG}\right)\ge0$.
Equation (\ref{eq:kzPTimpl}) can be solved explicitly as
\beq
k_z\left(\omega\right)=k\sqrt{\frac{\varepsilon_L\left(\omega\right)\varepsilon_G\left(\omega\right)}{\varepsilon_L\left(\omega\right)+\varepsilon_G\left(\omega\right)}},
\label{eq:kzPT}
\eeq
from which it appears rather evident the aforementioned formal analogy with the typical SPP dispersion law \cite{Maier:2007kw}. However, we stress that there are fundamental differences between the two phenomena, as the propagation mechanism in the GL bi-layer relies on permittivities with {\em positive} real-parts and {\em opposite-signed} imaginary parts. From the physical viewpoint, the  
propagation is sustained by a transverse (i.e., $x$-directed) power flow from the gain to loss region.

Especially interesting in this context is the $\mathcal{PT}$-symmetry condition (\ref{eq:PTS}), for which the dispersion law in (\ref{eq:kzPT}) assumes the form 
\beq
k_z\left(\omega_0\right)=k_0\sqrt{\frac{{\varepsilon'}_0^2+{\varepsilon''_0}^2}{2\varepsilon'_0}},
\label{eq:kz0}
\eeq
thereby yielding an {\em inherently-real} propagation constant. We highlight that, in principle, the above propagation mechanism can be sustained in the presence of {\em arbitrarily small} (but non-zero) gain/loss levels. Nevertheless, moderate to high levels of gain/loss are required for effective field confinement along the transverse ($x$) direction. Therefore, looking at the material dispersion in Fig. \ref{Figure2}, we can anticipate the existence of this type of surface-modes at any frequency. However, the field confinement is expected to be maximal at the center frequency $\omega_0$, and rapidly deteriorating when departing from there.

%%%%%%%%%%%%%%%%%%%%%%%%%%%%%%%%%%%%%%%%%%%%%%%%%%%%%%%%%%%%%%%%%%
\section{Representative Results}
%%%%%%%%%%%%%%%%%%%%%%%%%%%%%%%%%%%%%%%%%%%%%%%%%%%%%%%%%%%%%%%%%%
\label{Sec:Results}

%############################################################
%                Figure3
%
\begin{figure}
\begin{center}
\includegraphics [width=8cm]{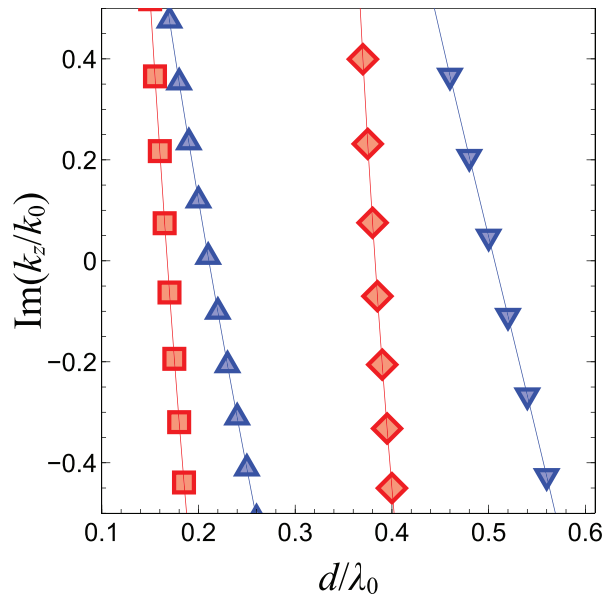}% Here is how to import EPS art
\end{center}
\caption{(Color online) Parameters as in Fig. \ref{Figure2}. Imaginary part of the propagation constant as a function of the gain-layer electrical thickness, for the four lowest-order modes at the center frequency $\omega_0$. Red squares and diamonds identify TE odd and even modes, respectively. Blue up- and down-triangles identify TM odd and even modes, respectively.}
\label{Figure3}
\end{figure}
%############################################################

%-----------------------------------------------------------------
\subsection{Generalities and Observables}
%-----------------------------------------------------------------

We move on to studying the modal solutions of the dispersion equations for the structures of interest. While the {\em algebraic} dispersion equation pertaining to the GL bi-layer can be readily solved in closed form [as shown in (\ref{eq:kzPT})], those pertaining to the TM and TE modes of the LGL three-layer [(\ref{eq:TM}) and (\ref{eq:TE}), with (\ref{eq:kxLG})] are of {\em transcendental} form, and need to be solved numerically in the complex $k_z$-plane (see Appendix \ref{Sec:AppendixB} for details).
%############################################################
%                Figure4
%
\begin{figure*}
\begin{center}
\includegraphics [width=16cm]{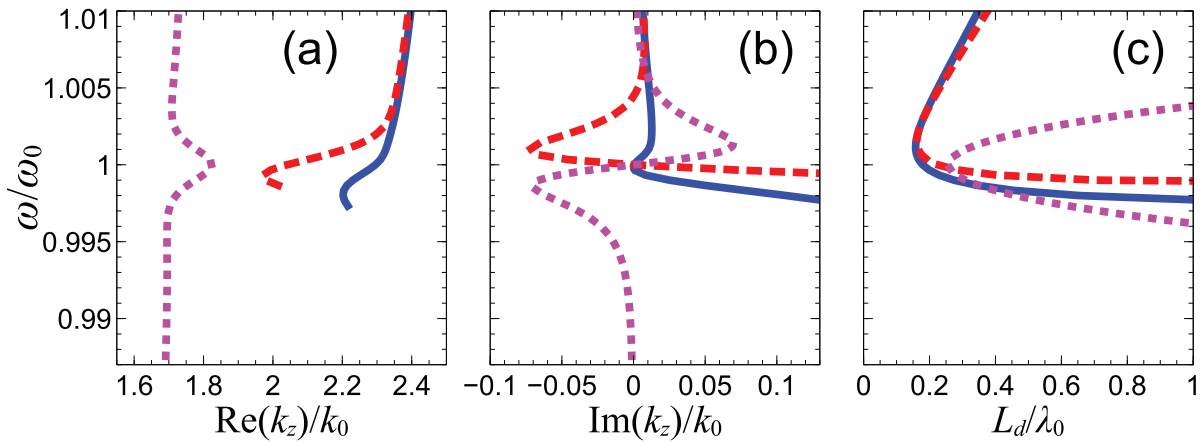}% Here is how to import EPS art
\end{center}
\caption{(Color online) Parameters as in Fig. \ref{Figure2}. (a), (b), (c) Real and imaginary part of the propagation constant $k_z$, and decay length $L_d$, respectively, as a function of frequency, for the TE ($d=0.168\lambda_0$, red-dashed curves) and TM ($d=0.211\lambda_0$, blue-solid curves) odd modes supported by the LGL three-layer. The gain-layer thickness $d$ is chosen so as to attain a purely real propagation constant at the center frequency $\omega_0$ (cf. Fig. \ref{Figure3}). Also shown (magenta-dotted curves) are the results pertaining to the TM mode supported by the GL bi-layer.}
\label{Figure4}
\end{figure*}
%############################################################

In what follows, results will be illustrated in terms of conventional dispersion (Brilluoin) diagrams, i.e., normalized frequency ($\omega/\omega_0$) as a function of real and imaginary parts of the normalized propagation constant $k_z/k_0$. It is important to highlight that, different from the plasmonic case, the presence of gain in our configurations allows either exponentially {\em decaying} [$\mbox{Im}\left(k_z\right)>0$] or {\em growing} [$\mbox{Im}\left(k_z\right)<0$] solutions. While the amplification capabilities may seem intriguing, it is also evident that on suitably long propagation distances the gain saturation would inevitably be reached, thereby driving the system away from the linear regime assumed in our study. Therefore, as for the plasmonic case, we are especially interested in modes with {\em quasi-real} propagation constants, i.e., $\mbox{Im}\left(k_z\right)\approx 0$.

Another meaningful observable, to quantitative assess the wave confinement along the transverse ($x$) direction, is the {\em decay length} \cite{Maier:2007kw}
\beq
L_d=\frac{1}{\left|\mbox{Im}\left(k_{xL}\right)\right|}.
\label{eq:Ld}
\eeq

%-----------------------------------------------------------------
\subsection{Critical Thickness}
%-----------------------------------------------------------------
\label{Sec:CT}
As previously discussed in Sec. \ref{Sec:GL} above, in view of the $\mathcal{PT}$-symmetry condition (\ref{eq:PTS}), the GL bi-layer inherently supports a TM surface mode with {\em purely real} propagation constant at the center frequency $\omega_0$ [see (\ref{eq:kz0})]. Conversely, the TM and TE dispersion equations in (\ref{eq:TM}) and (\ref{eq:TE}) generally yield {\em complex-valued} propagation constants. For a meaningful comparison, it makes sense to preliminary investigate the possibility to attain real-valued propagation constants in these cases too.

Figure \ref{Figure3} shows, for the lowest-order TM and TE (even and odd) modes, the imaginary part of the propagation constant at the center frequency $\omega_0$, as a function of the gain-layer electrical thickness. As it can be observed, for the assumed parameters, all curves are monotonically decreasing and pass through zero. In other words, there exist critical values of the electrical thickness (different for the four modes) that ensure a purely real propagation constant at $\omega_0$. In particular, these values are smaller (by roughly a factor two) for the odd modes, and the smallest possible ($d=0.168\lambda_0$) is observed for the TE odd mode.
Such value represents the {\em minimum} allowed gain-layer thickness for the structure to support a non-attenuating (and non-amplifying) mode at $\omega_0$, and therefore sets a limit for the structure miniaturization. To give an idea of the order of magnitude, for the assumed material parameters and dispersion (consistent with a gain medium based on CdSe/ZnS core-shell quantum-dots \cite{Campbell:2012cj} at $\lambda_0=560$nm), such minimum thickness would be around 94nm.

In the following studies, we disregard the even modes, and focus on the lowest-order TE and TM odd modes, which can operate with {\em electrically thinner} gain-medium layers.

%-----------------------------------------------------------------
\subsection{Study of TE and TM Odd Modes}
%-----------------------------------------------------------------
Figure \ref{Figure4} shows the results pertaining to the lowest-order TE and TM odd modes supported by our LGL three-layer, with material dispersion as in Fig. \ref{Figure2} and electrical thicknesses chosen according to Fig. \ref{Figure3}
(see the discussion in Sec. \ref{Sec:CT} above). More specifically, Figs. \ref{Figure4}(a) and \ref{Figure4}(b) show the dispersion diagrams (real and imaginary, respectively), while Fig. \ref{Figure4}(c) shows the decay length (\ref{eq:Ld}).
Also shown, for comparison, are the results pertaining to the TM surface-mode supported by the GL bi-layer, with same parameters.

A few observations are in order. As qualitatively anticipated (see the discussion in Sec. \ref{eq:DE} above), unlike the TM surface-mode of the GL bi-layer, the LGL modes exhibit a ``cut-off'' frequency very close to the center frequency ($\sim 0.997\omega_0$ for the TM-odd, and $\sim 0.999\omega_0$ for the TE-odd), below which no proper solutions [$\mbox{Im}\left(k_{xL}\right)\ge 0$] are found. Moreover, as expectable in view of the parameter choices, all three modes exhibit purely real propagation constants at $\omega_0$. At this frequency, the decay length is $L_d\approx 0.169\lambda_0$ for the modes of the LGL three-layer, and $L_d=0.259\lambda_0$ for the GL bi-layer. We emphasize that at $\omega_0$ there is exactly {\em zero-contrast} in the permittivity real parts (see Fig. \ref{Figure2}), and hence the observed wave confinement and guiding effects are entirely attributable to the non-Hermiticity. For small departures from $\omega_0$, the $k_z$ imaginary part becomes nonzero and the decay length increases, for all three modes. However, it is interesting to note that, for the TM-odd mode, the $k_z$ imaginary part is always positive, and maintains moderately small values ($\lesssim 0.01k_0$) for $\omega\gtrsim \omega_0$.
On the other hand, for the TE-odd and the GL bi-layer modes, the $k_z$ imaginary part can be either positive or negative, and significantly larger (up to a factor $\sim 7$, in absolute value). From Fig. \ref{Figure4}(c), it is also interesting to notice that, for $\omega\gtrsim\omega_0$, the decay length in the GL bi-layer case increases very rapidly, and becomes wavelength-sized for frequency variations as small as $\sim 0.4\%$. For the LGL three-layer, such increase (comparable for the TE and TM modes) is less pronounced. However, as expectable, the decay length diverges when approaching the cut-off frequency. As qualitatively anticipated (see the discussion in Sec. \ref{Sec:CT} above), for higher frequencies, the results asymptotically tend to those of a standard weakly-guiding dielectric waveguide.

%############################################################
%                Figure5
%
\begin{figure}
\begin{center}
\includegraphics [width=9cm]{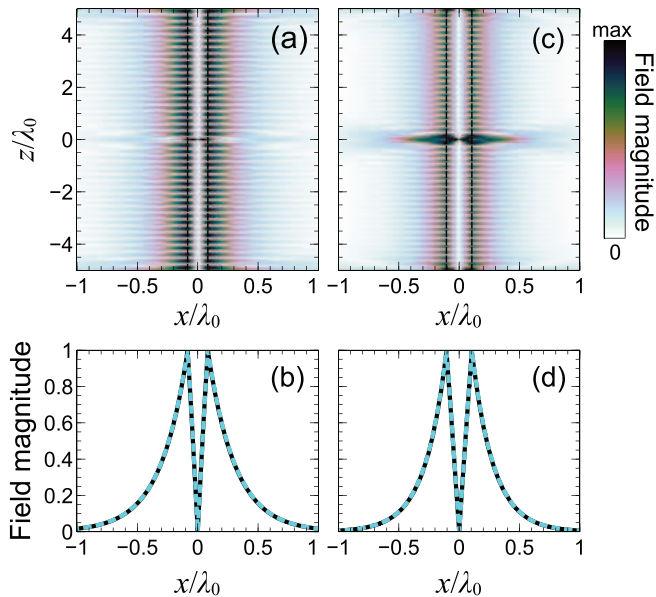}% Here is how to import EPS art
\end{center}
\caption{(Color online) Parameters as in Fig. \ref{Figure4}. (a) Finite-element-computed electric-field magnitude map ($|E_y|$) at the center frequency $\omega_0$ pertaining to the TE odd mode in an LGL three-layer of $10\lambda_0$-size along the $z$-direction (the gain layer is delimited by a black-dashed rectangle). The structure is excited by an electric line source located at $x=z=0$, and the false-color-scale is suitably saturated for better visualization. (b) Transverse cut (black-solid curve) at $z=2\lambda_0$, compared with analytical prediction (cyan-dashed curve). Fields are normalized with respect to their maxima.
(c), (d) Corresponding results for TM odd mode. In this case, the magnetic field ($|H_y|$) is displayed, and a magnetic line source located at $x=z=0$ is considered in the finite-element simulations.}
\label{Figure5}
\end{figure}
%############################################################

At a first glance of the above dispersion diagrams, one feature that may appear puzzling is the presence of {\em infinite-derivative} points, which seem to imply infinite values of the group velocity. This seeming physical inconsistency (which is also observed in plasmonic systems \cite{Dionne:2005pl}) can be lifted by recalling that the conventional association between group velocity and $d\omega/d k_z$ ceases to be meaningful in the presence of {\em anomalous dispersion} (cf. Fig. \ref{Figure2}), and more sophisticated approximations are needed (see, e.g., \cite{Tanaka:1989do}).
%############################################################
%                Figure6
%
\begin{figure}
\begin{center}
\includegraphics [width=9cm]{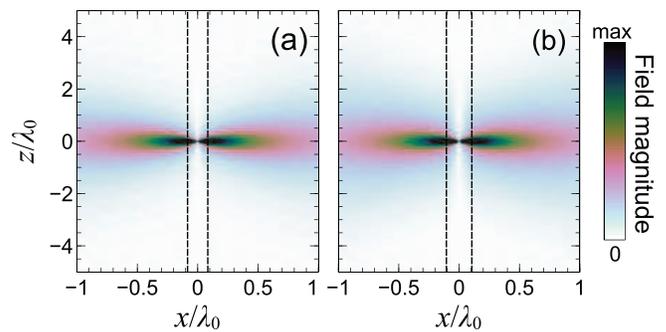}% Here is how to import EPS art
\end{center}
\caption{(Color online) (a), (b) As in Figs. \ref{Figure5}(a) and \ref{Figure5}(c), respectively, but at $\omega=0.995\omega_0$.}
\label{Figure6}
\end{figure}
%############################################################
Another seemingly counterintuitive issue is the above mentioned infinite growth of the decay length when approaching the cut-off frequency, which seems to imply unattenduated propagation in the lossy halfspaces. In fact, it is well known \cite{Nabulsi:1992dr} that this effect is a consequence of the unrealistic assumption of a traveling-wave $\sim\exp\left(i k_z z\right)$ propagation  with complex-valued propagation constant  over an {\em infinite} extent. When realistically limited to an aperture of finite extent, the enhanced penetration effect is in fact limited to the near-field region \cite{Nabulsi:1992dr}.
%############################################################
%                Figure7
%
\begin{figure}
\begin{center}
\includegraphics [width=9cm]{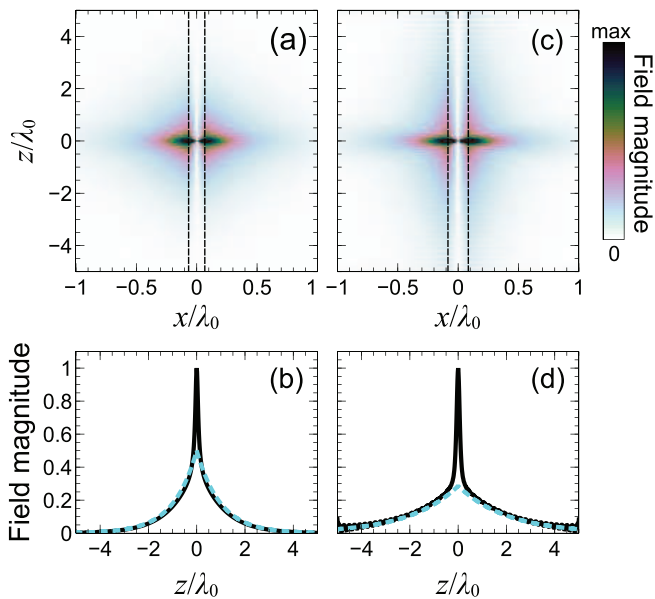}% Here is how to import EPS art
\end{center}
\caption{(Color online) (a), (c) As in Figs. \ref{Figure5}(a) and \ref{Figure5}(c), respectively, but with gain-layer thickness reduced by 20\% ($d=0.134\lambda_0$ and $d=0.169\lambda_0$, respectively). (b), (d) Corresponding longitudinal cuts (black-solid curves) at the gain-loss interface $x=d/2$, normalized with respect to their maxima. Also shown (cyan-dashed curves) as references are the exponentially-decay trends pertaining to the theoretical propagation constants $k_z=\pm(2.059+0.163i)k_0$ and $k_z=\pm (2.323+0.079i)k_0$, respectively.}
\label{Figure7}
\end{figure}
%############################################################

As an independent validation, to ascertain the physical character and actual excitability of the above studied modes, Figs. \ref{Figure5}(a) and \ref{Figure5}(c) show the finite-element computed (see Appendix \ref{Sec:AppendixA} for details) field maps for the TE and TM odd modes, respectively, pertaining to a finite-size (along $z$) structure excited via an (electric and magnetic, respectively) line source at $\omega_0$. The presence of transversely-confined modes propagating without attenuation/amplification is clearly observable, with a visible standing-wave pattern attributable to the structure truncation along the $z$-direction. Figures \ref{Figure5}(b) and \ref{Figure5}(d) show two transverse ($x$) cuts, compared with the theoretical predictions (see Appendix \ref{Sec:AppendixB}). As typical of odd modes, and in excellent agreement with the theoretical predictions, the fields are peaked at the gain-loss interfaces, vanish at the center of the gain layer, and decay exponentially in the loss regions.

For the same parameter configuration, Fig. \ref{Figure6} shows the field maps computed at $\omega=0.995 \omega_0$, i.e., slightly below the above mentioned cut-off frequencies. As it can be observed, in this case the line-source excitations cannot couple with guided modes, and essentially radiate in the lossy regions (see the discussion at the beginning of Sec. \ref{eq:DE} above).

As a further instructive example, Fig. \ref{Figure7} shows the results obtained by reducing of 20\% the gain-layer thickness. From the field maps [Figs. \ref{Figure7}(a) and \ref{Figure7}(c)], we qualitatively observe the excitation of {\em damped} modes. For a more quantitative assessment, Figs. \ref{Figure7}(b) and \ref{Figure7}(d) shows some relevant longitudinal cuts at the gain-loss interface $x=d/2$, from which an asymptotic exponential decay can be observed, in fairly good agreement with the theoretically-predicted {\em complex-valued} propagation constants.

%############################################################
%                Figure8
%
\begin{figure*}
\begin{center}
\includegraphics [width=16cm]{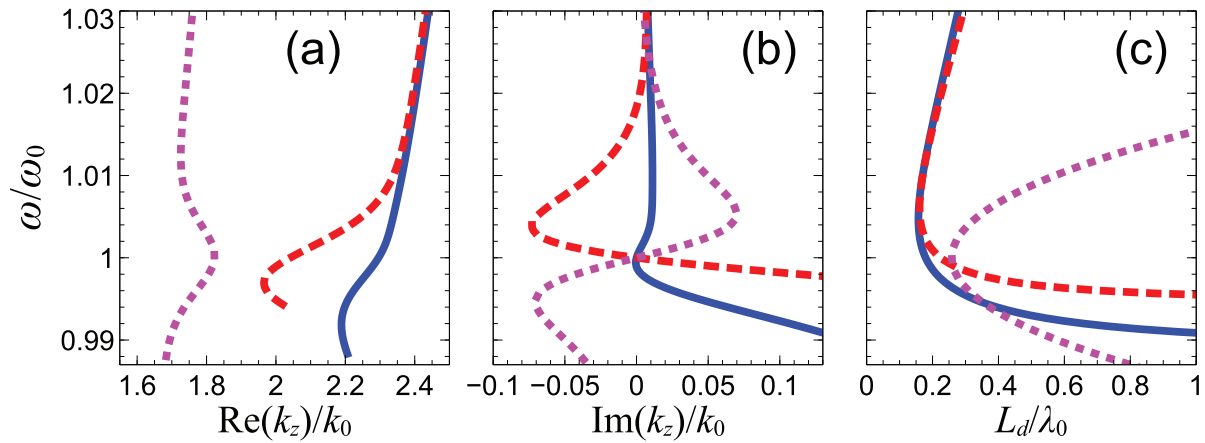}% Here is how to import EPS art
\end{center}
\caption{(Color online) As in Fig. \ref{Figure4}, but assuming $\Gamma=1.809\cdot10^{-2}\omega_0$ in (\ref{eq:Lorentz}), with all other parameters unchanged.}
\label{Figure8}
\end{figure*}
%############################################################
%-----------------------------------------------------------------
\subsection{Effects of Material Dispersion and Gain/Loss Level}
%-----------------------------------------------------------------
It is apparent from the above results that the dispersion properties of the modes are strongly tied with the material dispersion and parameters of the gain and loss media. Intuitively, broadening the Lorentzian lineshape in Fig. \ref{Figure2} enlarges the frequency range where  the non-Hermiticity-induced wave confinement and guiding are effective, and viceversa. Moreover, higher values of the peak gain/loss level yield stronger confinement effects, and viceversa.
To quantitatively elucidate these aspects, we study the effects induced by varying the dampening parameter $\Gamma$ and the peak gain/loss level $\varepsilon''_0$ in (\ref{eq:Lorentz}).

Figure \ref{Figure8} shows (in the same format as in Fig. \ref{Figure4}) the results obtained by increasing by a factor four the value of $\Gamma$ in (\ref{eq:Lorentz}), with all other parameters unchanged (i.e., broadening the lineshape while maintaining the same peak gain/loss level). We note that there is no need to re-calculate the critical values of the gain-layer thickness, as these depend solely on the permittivity values at $\omega_0$, which are not affected by $\Gamma$. 
Although the diagram look qualitatively similar to those in the previous example (cf. Fig. \ref{Figure4}), a spectral broadening of all features is observed. For the LGL three-layer modes, the cut-off frequencies  are now slightly lower ($0.994\omega_0$ for TE, and $0.988\omega_0$ for TM), and the range $\omega\gtrsim\omega_0$ where the confinement remains effective is moderately broader. However, as expectable, the minimum decay length is identical as the previous example (cf. Fig. \ref{Figure4}). Moreover, also in this case, the TM-odd mode maintains a small (and almost always positive) imaginary part of the propagation constant.
%############################################################
%                Figure9
%
\begin{figure}
\begin{center}
\includegraphics [width=8cm]{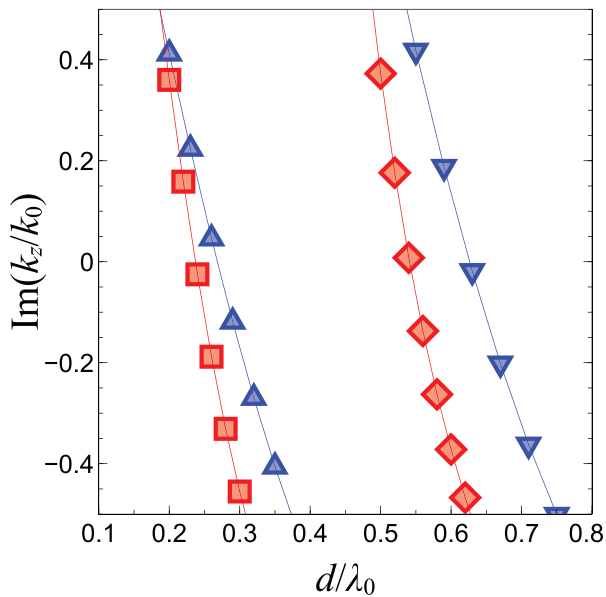}% Here is how to import EPS art
\end{center}
\caption{(Color online) As in Fig. \ref{Figure3}, but assuming $\varepsilon''_0=1.055$ in (\ref{eq:Lorentz}), with all other parameters unchanged.}
\label{Figure9}
\end{figure}
%############################################################

%############################################################
%                Figure10
%
\begin{figure*}
\begin{center}
\includegraphics [width=16cm]{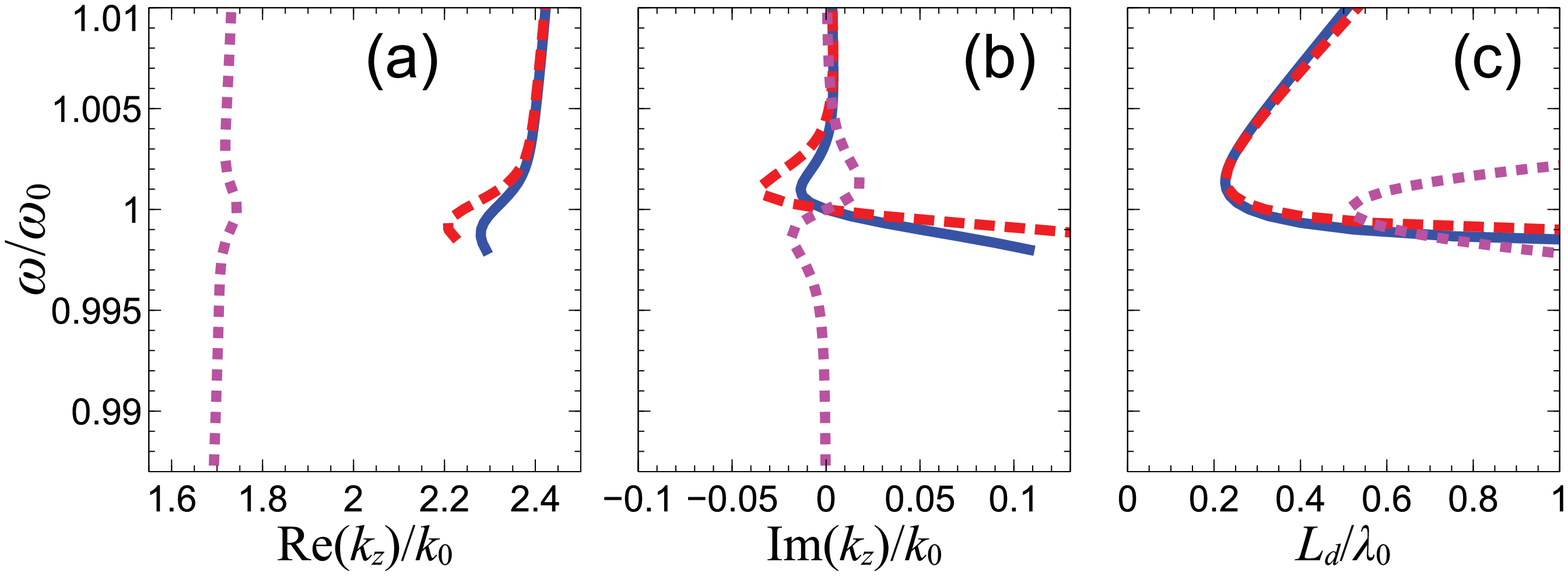}% Here is how to import EPS art
\end{center}
\caption{(Color online) As in Fig. \ref{Figure4}, but assuming $\varepsilon''_0=1.055$ in (\ref{eq:Lorentz}), with all other parameters unchanged. The gain-layer thicknesses in the LGL three-layer are now $d=0.237\lambda_0$ for the TE mode, and $0.268\lambda_0 $ for the TM mode (cf. Fig. \ref{Figure9}).}
\label{Figure10}
\end{figure*}
%############################################################

As a further example, we now consider a decrease by a factor two of $\varepsilon''_0$ in (\ref{eq:Lorentz}), with all other parameters identical as in Fig. \ref{Figure2}. This increases the peak gain/loss level, without changing the lineshape width.
In this case, the critical gain-layer thickness values can no longer be obtained from Fig. \ref{Figure3}, and need to be re-calculated. By repeating the same procedure as before, with the new value of $\varepsilon''_0$, we obtain Fig. \ref{Figure9}, from which we observe that the new critical thickness values are $d=0.237\lambda_0$ for the TE-odd mode, and $0.268\lambda_0 $ for the TM-odd mode. By comparison with the previous example, the minimum critical thickness has increased by a factor $\sim 1.4$.
Figure \ref{Figure10} shows the corresponding dispersion and decay-length diagrams. Once again, all previous qualitative observations still hold, and the most visible effect is a sensible increase of the decay length. In particular, at $\omega_0$, we observe $L_d=0.229\lambda_0$ for the LGL three-layer (i.e., a factor $\sim 1.36$ increase, by comparison with the previous examples) and $L_d=0.518\lambda_0$ for the GL bi-layer (i.e., a factor-two increase). Moreover, the TM-odd mode now also exhibits negative values of the propagation-constant imaginary part, even though they are sensibly smaller ($\lesssim 0.01k_0$ in absolute value) than those observed for the other modes.

The above examples illustrate how the material dispersion lineshape and parameters affect the wave confinement and guiding in the structures of interest. Clearly, different combinations of these effects are possible. Although, for simplicity of illustration, in our prototype study we have considered the same type of dispersion law and identical parameters for the gain and loss media, different choices are possible and additional degrees of freedom can be introduced in the models.

%-----------------------------------------------------------------
\subsection{Some Remarks}
%-----------------------------------------------------------------
\label{Sec:Remarks}

The above results indicate that LGL three-layers may provide some interesting advantages by comparison with the GL bi-layer studied in \cite{Ctyroky:2010kl,Savoia:2015}. In particular, the lowest-order TM-odd mode seems to offer the best tradeoff among small critical thickness of the gain-layer, good transverse field confinement, and quasi-real propagation constant (for $\omega\gtrsim\omega_0$).

It makes sense to wonder how the above results would be affected by the inevitable truncation (along the $x$-direction) of the lossy half-spaces. Although a numerical study is beyond the scope of this prototype study, these truncation effects are qualitatively similar to those observed for the GL bi-layer \cite{Ctyroky:2010kl,Savoia:2015}. In essence, assuming that the truncated structure is embedded in vacuum, as long as the thickness of these truncated layers is reasonably sized by comparison with the decay length, the wave confinement and guiding mechanism remains effective, and our model above provides a good approximation of the dispersion characteristics. For decreasing values of such thickness, the mechanism eventually ceases to be effective, and there is a smooth transition to a {\em leaky} regime, where energy is radiated in the outer vacuum space. Basically, a given truncation size determines a {\em threshold} value for the peak gain/loss level, below which the wave confinement and guiding mechanism cannot be sustained.

%%%%%%%%%%%%%%%%%%%%%%%%%%%%%%%%%%%%%%%%%%%%%%%%%%%%%%%%%%%%%%%%%%
\section{Conclusions}
%%%%%%%%%%%%%%%%%%%%%%%%%%%%%%%%%%%%%%%%%%%%%%%%%%%%%%%%%%%%%%%%%%
\label{Sec:Conclusions}

In summary, we have revisited  the wave confinement and guiding mechanism that can be entirely induced by non-Hermiticity
in LGL three-layers.
By using realistic models for the gain and loss materials, we have studied the dispersion and confinement properties of the lowest-order TE and TM modes, and compared them with the previously studied TM surface modes in GL bi-layers, highlighting similarities, differences, and possible superior characteristics. We have also illustrated the effects of material dispersion and parameters.

Our studies indicate that  LGL three-layers (in particular, TM-odd modes) tend to exhibit more desirable properties than those observed in GL bi-layers. Overall, the wave-confinement capabilities are essentially determined by the gain/loss level. Although the intended focus of this prototype study was on the illustration of the phenomenology, rather than the engineering of a specific application, we did assume realistic dispersion models and parameters for the material constituents. In particular, considering a commercially available gain medium based on CdSe/ZnS core-shell quantum-dots \cite{Campbell:2012cj} operating at $\lambda_0=560$nm), a layer of thickness as small as $\sim 120$nm ($0.211\lambda_0$) should be able to effectively sustain the waveguiding mechanism, with sub-wavelength transverse confinement (decay length $L_d\sim 0.17\lambda_0$).

The above results provide additional perspectives in the emerging framework of non-Hermitian optics, with potentially intriguing applications to novel optical devices and reconfigurable nanophotonics platforms. Accordingly, as a possible follow-up, we are currently studying the potential implications of these concepts in the design of non-Hermitian resonant particles in cylindrical and spherical core/multi-shell geometries. Also of interest are {\em full-wave} revisits of non-Hermitian multilayered configurations (e.g., LGLGL) that have been insofar investigated only within the (linear or nonlinear) paraxial regime (see, e.g., \cite{Malomed:2014ss} and references therein).

\appendix 

%%%%%%%%%%%%%%%%%%%%%%%%%%%%%%%%%%%%%%%%%%%%%%%%%%%%%%%%%%%%%%%%%%
\section{Details on Modal Solutions}
%%%%%%%%%%%%%%%%%%%%%%%%%%%%%%%%%%%%%%%%%%%%%%%%%%%%%%%%%%%%%%%%%%
\label{Sec:AppendixA}
In our formulation, TM and TE modes are characterized by $(E_x,E_z,H_y)$ and $(E_y,H_x,H_z)$ nonzero field components, respectively. For the LGL three-layer in Fig. \ref{Figure1}(a), guided-mode solutions can be generically expressed in terms of the $y$-directed fields as 
\begin{widetext}
\beq
\left\{E_y,H_y\right\}\left(x,z\right)=\exp\left(ik_z z\right)\left\{
\begin{array}{lll}
C_1 \exp\left(-ik_{xL}x\right),\hspace*{3.cm}x<-d/2,\\
C_2\exp\left(-ik_{xG}x\right)+C_3\exp\left(ik_{xG}x\right),~~\left|x\right|<d/2,\\
C_4 \exp\left(ik_{xL}x\right),\hspace*{3.3cm}x>d/2,\\
\end{array}
\right.
\label{eq:TETMf}
\eeq
\end{widetext}
with $k_{xL}$ and $k_{xG}$ defined as in (\ref{eq:kxLG}), and $C_j$, $j = 1,...,4$ denoting unknown expansion coefficients
to be calculated by enforcing the continuity of the electric and magnetic tangential fields at the interfaces $x=\pm d/2$.
From (\ref{eq:TETMf}), the remaining tangential field components can be straightforwardly derived via the relevant (source-free) Maxwell's curl equations. One ends up with a $4\times 4$ homogeneous linear system of equations, from which the dispersion equations follow by enforcing nontrivial solutions (i.e., zero determinant). In view of the inherent symmetry of the structure, these dispersion equations can be split into pairs (see, e.g., Sec. 2.3 in \cite{Maier:2007kw}), yielding the final results in (\ref{eq:TM}) and (\ref{eq:TE}). For instance, (\ref{eq:TM_even}) describes TM modes of {\em even} vector parity ($E_z$ even, $H_y$ and $E_x$ odd), while (\ref{eq:TM_odd}) pertains to TM modes of {\em odd} vector parity ($E_z$ odd, $H_y$ and $E_x$ even). Similar considerations hold for the TE even and odd modes in (\ref{eq:TE_even}) and (\ref{eq:TE_odd}), respectively.

As previously mentioned, the modal solutions for the GL bi-layer formally resemble the structure of SPPs. In this case, only TM modes can be supported, and a solution exponentially bound at the gain-loss interface can be expressed as
\beq
H_y\left(x,z\right)=C\exp\left(ik_z z\right)\left\{
\begin{array}{ll}
\exp\left(ik_{xG}x\right),~~x<0,\\
\exp\left(ik_{xL}x\right),~~x>0,\\
\end{array}
\right.
\label{eq:Hy}
\eeq
with $C$ denoting a constant, and the continuity at the interface $x=0$ already enforced. Unlike the LGL case, the branch-cut for $k_{xG}$ in (\ref{eq:kxG}) needs to be chosen consistently as $\mbox{Im}\left(k_{zG}\right)\ge0$, so as to satisfy the decay-at-infinity condition (see also the discussion in Sec. II-B of \cite{Savoia:2015}). The dispersion equation in (\ref{eq:kzPTimpl}) readily follows by enforcing the continuity at $x=0$ of the tangential electric field [derived from (\ref{eq:Hy}) and the relevant Maxwell's curl equation]. 
The explicit form in (\ref{eq:kzPT}) is obtained by squaring and solving with respect to $k_z$. However, care should be exerted since the squaring may actually introduce some spurious solutions that do not satisfy (\ref{eq:kzPTimpl}).

Finally, we note that, as for the plasmonic case \cite{Maier:2007kw}, the dispersion equation for the GL bi-layer in (\ref{eq:kzPTimpl}) could also be directly derived as the infinite-thickness ($d\rightarrow\infty$) limit  of the LGL three-layer TM counterparts (\ref{eq:TM}), with consistent choice of the branch-cuts.

%%%%%%%%%%%%%%%%%%%%%%%%%%%%%%%%%%%%%%%%%%%%%%%%%%%%%%%%%%%%%%%%%%
\section{Details on Numerical Modeling and Simulations}
%%%%%%%%%%%%%%%%%%%%%%%%%%%%%%%%%%%%%%%%%%%%%%%%%%%%%%%%%%%%%%%%%%
\label{Sec:AppendixB}
The results in Figs. \ref{Figure3}, \ref{Figure4}, and \ref{Figure8}--\ref{Figure10} are obtained by numerically solving
the dispersion equations (\ref{eq:TM}) and (\ref{eq:TE}) [with (\ref{eq:kxLG}) and (\ref{eq:Lorentz})] in the complex $k_z$ plane. Our Python-based numerical algorithm relies on the inexact Newton method \cite{Kelley:1995im}, implemented in the {\tt scipy.optimize.root} routine available in the SciPy optimization library \cite{SciPy}. In particular, among the possible options, we select the Broyden's first Jacobian approximation. For a given value of $\omega$, the algorithm is initialized with starting points $k_{zs}$ finely sampled within the range $[0.05k_0, 3k_0]$. Among the (generally complex) calculated solutions, we disregard as improper (unphysical) those with $\mbox{Im}\left(k_{xL}\right)<0$, since they do not satisfy the decay-at-infinity condition.

The field maps and corresponding cuts in Figs. \ref{Figure5}--\ref{Figure7} are computed by means of the finite-element-based commercial software package COMSOL Multiphysics \cite{COMSOL:2015}. In particular, we utilize the RF module, and consider a computational domain $\sim9\lambda_0\times 10\lambda_0$ (only partially shown in Figs. \ref{Figure5}--\ref{Figure7} for better visualization), suitably excited by electric (for TE modes) and magnetic (for TM modes) line sources, and
truncated by perfectly-matched-layers in all directions. However, due to the presence of a the central gain layer, reflections from the truncated (along $z$) edges of the structure are unavoidable, and standing-wave patterns may appear in the field maps (see, e.g., Fig. \ref{Figure5}). 
The computational domain is discretized with a triangular mesh of adaptive size, resulting in about 5 million degrees of freedom. The MUMPS solver is utilized, with default parameters.

%\begin{thebibliography}{99}

%\bibliography{PRA_LGW-WG}

%\end{thebibliography}


\end{document}